\documentclass[10pt, paper=a4, UKenglish]{article}
\usepackage{graphicx}
\def\Title#1{\begin{center} {\Large #1 } \end{center}}
\def\Author#1{\begin{center}{ \sc #1} \end{center}}
\def\Address#1{\begin{center}{ \it #1} \end{center}}

\newcommand\pubblock{\rightline{\begin{tabular}{l} Proceedings of the CTD 2025\\ \pubnumber\\
         \pubdate  \end{tabular}}}

\newenvironment{Abstract}{\begin{quotation} \begin{center} 
             \large ABSTRACT \end{center}\bigskip 
      \begin{center}\begin{large}}{\end{large}\end{center} \end{quotation}}

\newenvironment{Presented}{\begin{quotation} \begin{center} 
             PRESENTED AT\end{center}\bigskip 
      \begin{center}\begin{large}}{\end{large}\end{center} \end{quotation}}

\def\Acknowledgements{\bigskip  \bigskip \begin{center} \begin{large}
      \bf ACKNOWLEDGEMENTS \end{large}\end{center}}

\def\beq{\begin{equation}}
\def\eeq#1{\label{#1}\end{equation}}
\def\eeqn{\end{equation}}

\def\beqa{\begin{eqnarray}}
\def\eeqa#1{\label{#1}\end{eqnarray}}
\def\eeqan{\end{eqnarray}}

\let\bar=\overbar

\def\Dslash{\not{\hbox{\kern-4pt $D$}}}
\def\dslash{\not{\hbox{\kern-2pt $\del$}}}

\def\msb{{\bar{\ssstyle M \kern -1pt S}}}

\usepackage{color}
\usepackage{lineno}
\usepackage{subfig}
\usepackage{hyperref}
\usepackage{xspace}
\usepackage{doi}

\newcommand\pubnumber{PROC-CTD2025-049}

\newcommand\pubdate{\today}

\newcommand{\conference}{Connecting the Dots Workshop (CTD 2025)\\
November 10-14, 2025}

\usepackage{fancyhdr}
\definecolor{mygrey}{RGB}{105,105,105}
\usepackage{amsmath}
\usepackage{cleveref}

\begin{document}


\large
\begin{titlepage}
\pubblock

\vfill
\Title{Mitigating Detector Ageing Effects with Graph-Based Multi-Modal Track Reconstruction at Belle~II}
\vfill

\Author{Lea Reuter$^{1}$, Tristan Brandes$^{1}$, Giacomo De Pietro$^{1,2}$ and Torben Ferber$^{1}$}
\Address{$^1$Insitute for Experimental Particle Physics (ETP), Karlsruhe Institute of Technology (KIT),
Karlsruhe, Germany\\
$^2$Scientific Computing Center (SCC), Karlsruhe Institute of Technology (KIT), Karlsruhe,
Germany}

\vfill

\newcommand{\belle}{Belle~II\xspace}
\newcommand{\cdc}{\ensuremath{\mathrm{CDC}}\xspace}
\newcommand{\svd}{\ensuremath{\mathrm{SVD}}\xspace}
\newcommand{\legendre}{\textit{Baseline Finder}\xspace}
\newcommand{\cat}{\textit{CAT Finder}\xspace}
\newcommand{\bat}{\textit{BAT Finder}\xspace}
\def \opengen {\ensuremath{\alpha^{3D}_{\mathrm{MC}}}\xspace}
\def \pgen {\ensuremath{p_{\mathrm{MC}}}\xspace}
\def \ptgen {\ensuremath{p_{\mathrm{t,gen}}}\xspace}
\def \thetagen {\ensuremath{\theta_{\mathrm{MC}}}\xspace}
\def \phigen {\ensuremath{\phi_{\mathrm{MC}}}\xspace}
\def \zgen {\ensuremath{z_{\mathrm{MC}}}\xspace}
\def \rhogenthreed {\ensuremath{r^{3D}_{\mathrm{MC}}}\xspace}
\def \pyg {\texttt{PyTorch Geometric}\xspace}
\def \pytorch {\texttt{PyTorch}\xspace}
\def \basf {\texttt{basf2}\xspace}
\def \layereff {\ensuremath{\varepsilon_{\text{layer,i}}}\xspace}

\begin{Abstract}
Large backgrounds that can lead to hardware failures and the degradation of detector gain impact the track finding in the \belle central drift chamber.
These conditions lead to spatially non-uniform and time-dependent inefficiencies,  which results in inactive regions and missing hits which challenges conventional tracking algorithms and necessitate the development of new track finding algorithms.
In this work, we evaluate the performance of our previously developed unified graph neural network (GNN) based track-finding algorithm under realistic long-term detector ageing conditions.
Track finding is formulated as a global relational clustering problem using object condensation, which enables the reconstruction of an unknown and variable number of tracks per event.
Using a realistic full detector simulation incorporating beam-induced backgrounds, detector noise, and measured detector ageing effects, we evaluate the tracking performance and compare it to the current \belle baseline reconstruction. 
We show that detector degradation can be treated as a domain shift in the observed hit patterns, rather than requiring a fundamentally new reconstruction strategy.
After retraining on degraded detector conditions, the GNN-based approach limits the absolute track efficiency loss for uniformly displaced muons to 14\%, compared to 28\% for the baseline tracking, while maintaining a track purity of 96\%. Under the same conditions, the baseline reconstruction achieves only 90\% track purity.
These results demonstrate that the unified GNN-based reconstruction provides increased robustness to irregular hit patterns and extended inactive regions, enabling stable tracking performance under long-term detector ageing at \belle.
\end{Abstract}

\vfill

\begin{Presented}
\conference
\end{Presented}
\vfill
\end{titlepage}
\def\thefootnote{\fnsymbol{footnote}}
\setcounter{footnote}{0}

\newcommand{\belle}{Belle~II\xspace}
\newcommand{\cdc}{\ensuremath{\mathrm{CDC}}\xspace}
\newcommand{\svd}{\ensuremath{\mathrm{SVD}}\xspace}
\newcommand{\legendre}{\textit{Baseline Finder}\xspace}
\newcommand{\cat}{\textit{CAT Finder}\xspace}
\newcommand{\bat}{\textit{BAT Finder}\xspace}
\def \opengen {\ensuremath{\alpha^{3D}_{\mathrm{MC}}}\xspace}
\def \pgen {\ensuremath{p_{\mathrm{MC}}}\xspace}
\def \ptgen {\ensuremath{p_{\mathrm{t,gen}}}\xspace}
\def \thetagen {\ensuremath{\theta_{\mathrm{MC}}}\xspace}
\def \phigen {\ensuremath{\phi_{\mathrm{MC}}}\xspace}
\def \zgen {\ensuremath{z_{\mathrm{MC}}}\xspace}
\def \rhogenthreed {\ensuremath{r^{3D}_{\mathrm{MC}}}\xspace}
\def \pyg {\texttt{PyTorch Geometric}\xspace}
\def \pytorch {\texttt{PyTorch}\xspace}
\def \basf {\texttt{basf2}\xspace}
\def \layereff {\ensuremath{\varepsilon_{\text{layer,i}}}\xspace}
\normalsize 


\section{Introduction}
\label{sec:introduction}

The operation of the SuperKEKB electron-positron collider at high instantaneous luminosity poses significant challenges for the \belle tracking system.
Beam-induced backgrounds arise from particle losses during machine operation, producing electromagnetic showers that generate high hit rates in the \belle detector. 
For the main tracking device, the Central Drift Chamber (\cdc), one primary source of hit occupancy is the combination of high beam currents and small bunch sizes~\cite{NATOCHII2023168550} necessary to achieve the design instantenous luminosity of $6 \times 10^{35}\,\text{cm}^{-2}\text{s}^{-1}$~\cite{Belle-II:2010dht}. 
These conditions enhance beam-gas scattering and intra-bunch scattering.
The detector occupancy and radiation dose is further increased due to collision processes with high cross section, such as radiative Bhabha scattering and two photon processes~\cite{Natochii:2022vcs,aihara2024belle}.
Although the currently achieved instantaneous luminosity remains approximately one order of magnitude below its design value, the hit occupancy and resulting operating currents in the \cdc have already increased substantially and approached acceptable limits for the current operation~\cite{BelleIIReport2025}.
The increased occupancy particularly affects the innermost layers of the \cdc, where accumulated charges of up to $\mathcal{O}(150\,\text{mC/cm})$ per wire have been observed~\cite{BelleIIReport2025}.
This leads to a progressive loss of gas gain, an increased risk of Malter-like currents~\cite{malter}, and a higher probability of short-term hardware failures of the readout boards~\cite{taniguchi2017central}, referred to as detector ageing.
\\
Conventional tracking algorithms assume that charged particles produce approximately continuous and geometrically ordered sequences of hits in each detector subsystem~\cite{MANKEL1999268,Mankel_2004,cmsCollaboration_2014,atlasCollaboration_2017}.
With increasing \cdc ageing, these assumptions no longer hold: reduced gas gain leads to random hit losses on an event-by-event basis, while disabled readout boards cannot record data in their respective detector regions on time scales of days to weeks. 
These effects produce a spatially non-uniform and time-dependent detector response, resulting in irregular and unpredictable degradation of hit patterns.
This behaviour is challenging for reconstruction approaches that rely on geometric continuity and stable detector coverage.
\\
\begin{figure}
    \centering
    \includegraphics[width=0.95\linewidth]{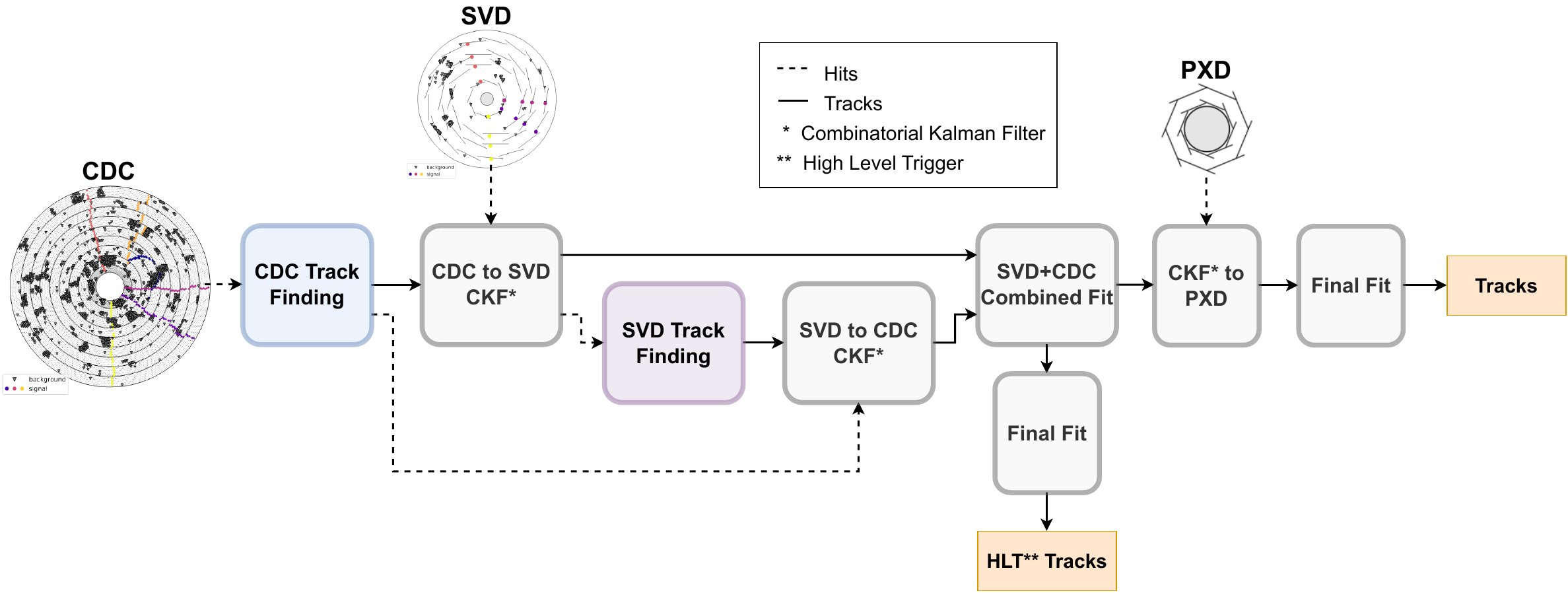}
    \caption{The \belle track finding chain for high level trigger tracks and tracks used in offline reconstruction. The blue box shows the CDC track finding algorithm, while the purple box indicates the SVD track finding algorithm. Final objects are highlighted in orange. The inputs of the three tracking detectors CDC, SVD, and PXD are given by one example event display, where the hits from signal particles are shown in colored circular markers and the beam-background is shown with grey triangular markers. See text for detail. Taken from \cite{acat}.}
    \label{fig:baseline_tracking}
\end{figure}
The Belle~II tracking system consists, from the beam-pipe outward, of the Pixel Detector (PXD), the Silicon Vertex Detector (\svd), and the Central Drift Chamber (\cdc)~\cite{Belle-II:2010dht}.
The baseline \belle reconstruction (\legendre)~\cite{BelleIITrackingGroup:2020hpx} is shown in \cref{fig:baseline_tracking}. 
It follows a staged approach in which track finding is performed sequentially in individual tracking subdetectors, starting from the \cdc-only track finding. The found track candidates are then extrapolated using a Combinatorial Kalman filter (CKF) to the \svd, where \svd hits are assigned to the track candidates. 
The \svd-only track finding is then performed on the remaining \svd hits, and the new track candidates are extrapolated to the \cdc to attach remaining \cdc hits. 
The track candidates are extrapolated to the innermost subdetector, the silicon PXD, and hits are attached to the track candidates.
Finally, these track candidates are passed to the standard Belle~II track fitting with the GENFIT2 Kalman filter based on Deterministic Annealing Filter~\cite{Hoppner:2009af,Rauch:2014wta,Bilka:2019ang, GENFIT_zenodo}.
For tracks used for the high level trigger (HLT) decisions, the extrapolation to the PXD is skipped.
While this current strategy  allows for some isolated missing hits, to suppress the reconstruction of fake tracks the \legendre employs a track quality criteria that removes tracks with extended gaps between consecutive hits.
Due to this track quality criteria, large numbers of missing hits due to the degraded \cdc lead to an abrupt loss of reconstruction efficiency.
\\
Graph Neural Networks (GNNs) proved to be a successful alternative track finding approach at \belle compared to the \legendre~\cite{cat_paper,acat}.
The GNN-based algorithm uses as input an unordered set of hits and learns which hits to connect in order to exchange information and form track candidates.
The advantage over the \legendre reconstruction chain is that it does not rely on explicit detector geometry or sequential hit patterns.
\\
In this work, we study realistic long-term \cdc ageing effects, including inefficiencies and disabled readout boards, and their impact on track reconstruction.
We investigate whether these effects can be addressed through data-driven adaptation of a GNN-based track finder.
This proceeding describes the detector ageing mechanisms in \cref{sec:ageing}, the GNN framework in \cref{sec:gnn}, the training strategy for degraded detector conditions in \cref{sec:training}, and the resulting tracking performance in \cref{sec:performance}, before concluding in \cref{sec:conclusion}.

\section{CDC Degradation in Belle~II}
\label{sec:ageing}

\begin{figure}
    \centering
    \subfloat[]{\includegraphics[width=0.45\linewidth]{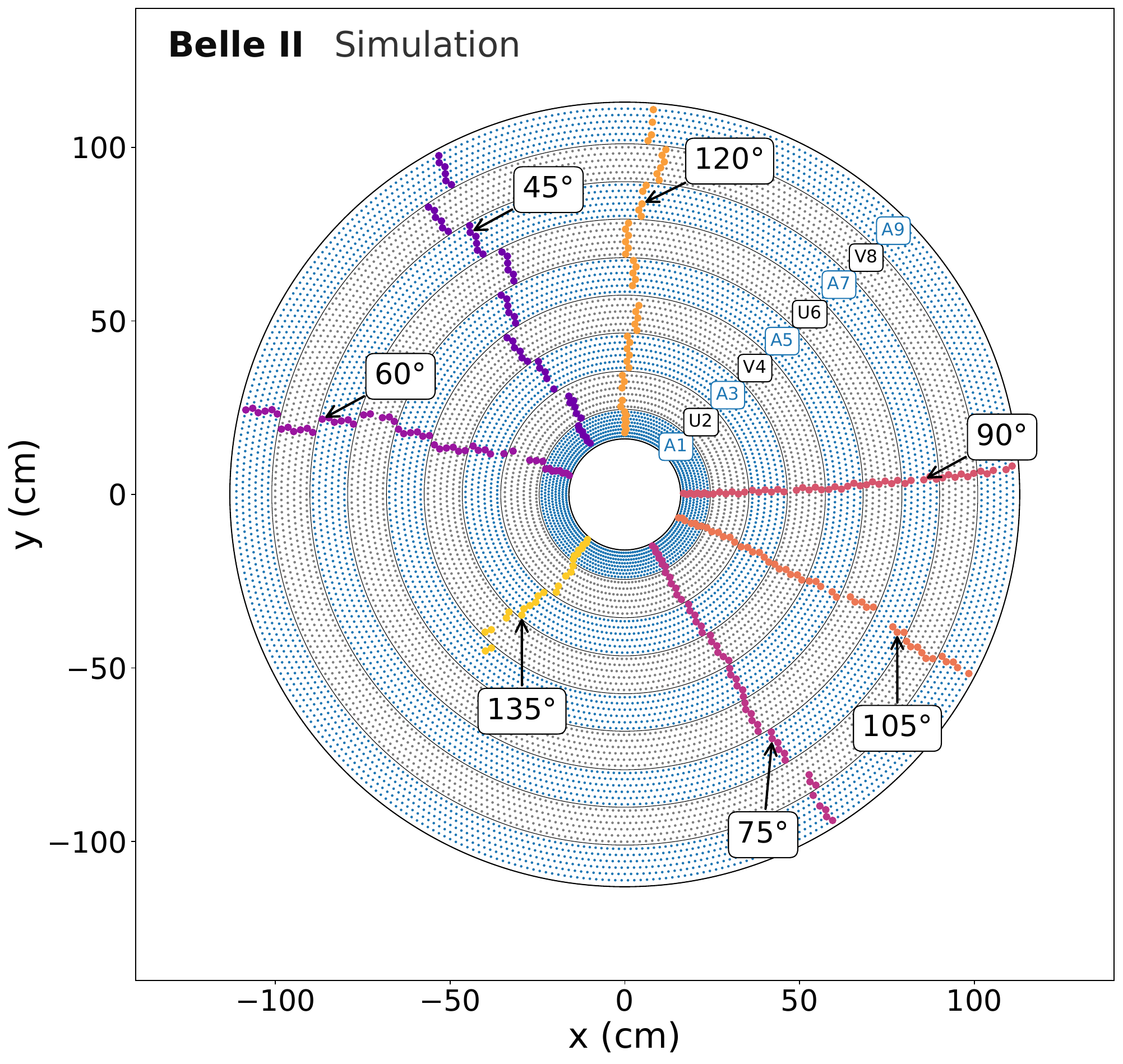}}
    \qquad
     \subfloat[]{\includegraphics[width=0.45\linewidth]{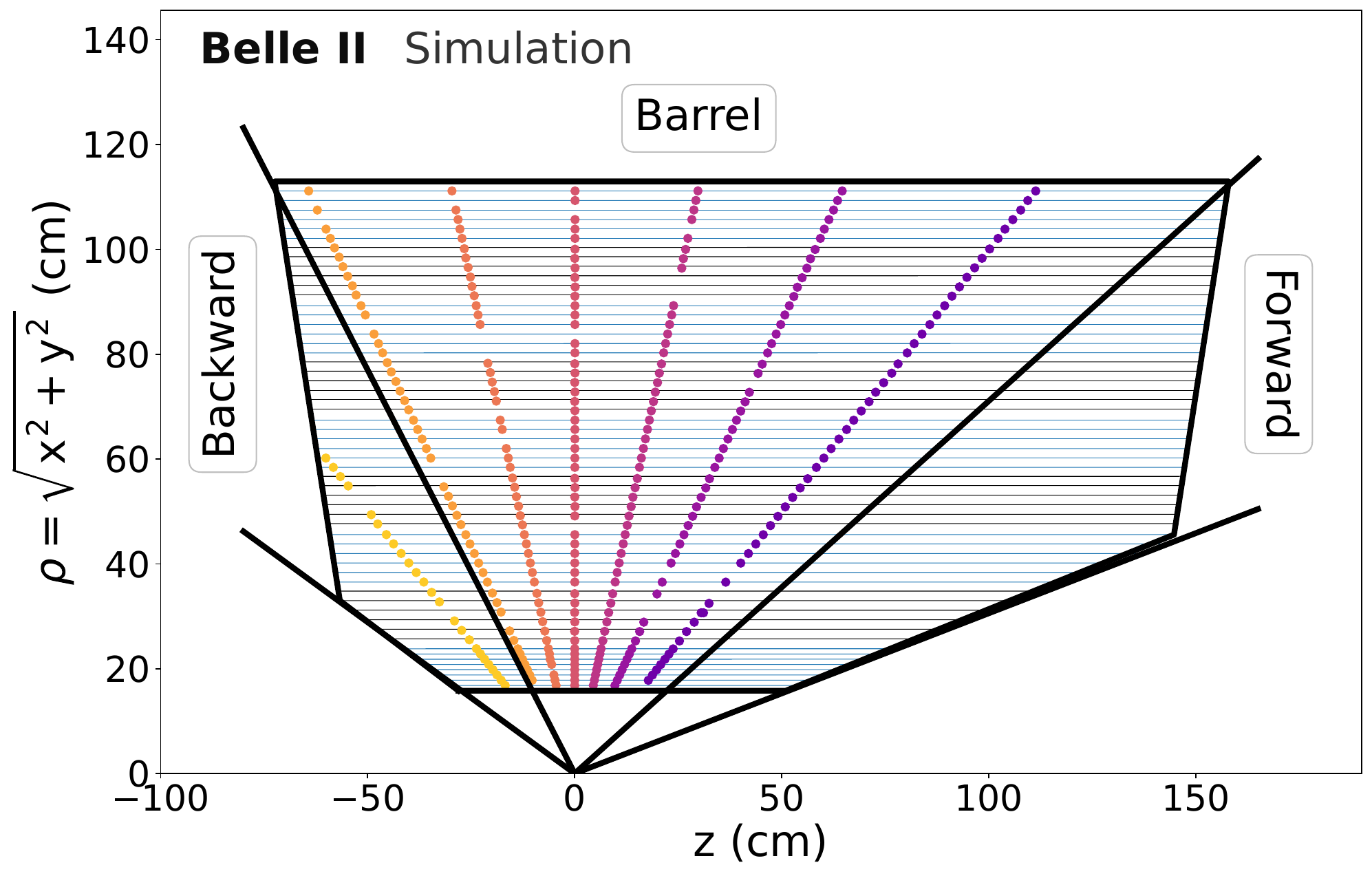}}
    \caption{Geometrical structure of the \cdc shown in $x-y$ view (a) and $z-\rho$ view with $\rho=\sqrt{x^2+y^2}$ in (b).
    Sense wires in axial layers are shown in blue, while sense wires in stereo layers are shown in black. 
    Coloured hits from signal particles at different polar angles $\theta$ illustrate the offset between axial and stereo layers used to infer the longitudinal track direction by the track finding algorithm.}
    \label{fig:cdc_geometry}
\end{figure}

The \cdc is a gaseous tracking detector with an outermost radius of 113\,cm, composed of more than 14\,000 sense wires arranged in alternating superlayers of axial and stereo wires~\cite{taniguchi2017central,dong2019calibration}, illustrated in \cref{fig:cdc_geometry}.
The \belle coordinate system is defined such that the positive $z$-axis is approximately aligned with the electron beam direction and corresponds to the solenoid axis. 
The $x$-axis points horizontally away from the detector, while the $y$-axis points vertically upward.
The polar angle $\theta$ is measured with respect to the $z$-axis, and the azimuthal angle $\phi$ is defined in the transverse plane.
Each measured hit in the \cdc provides spatial information in the plane perpendicular to the sense wire: for axial layers this corresponds to the \belle $x$-$y$ plane, while for stereo layers the measurement plane is rotated according to the wire skew angle.
The longitudinal coordinate along the wire is not measured, as signals are read out from only one end.
The CDC measures drift times relative to the sense wires, providing only two-dimensional information in the plane perpendicular to the wire.
The full three-dimensional hit position can be determined only after a track hypothesis is available, as the wire geometry and stereo angles must be combined with an assumed trajectory to resolve the longitudinal coordinate.
\\
The performance of the \cdc is sensitive to radiation exposure through high background-induced hit rates, and long-term charge accumulation on the sense wires~\cite{BelleIIReport2025}.
As SuperKEKB approaches higher luminosities, several degradation mechanisms increasingly affect the CDC and thus the tracking performance.

\subsubsection*{Accumulated charge and wire ageing}
Sustained operation at high occupancy leads to charge accumulation on the sense wires, modifying the effective gas gain and gradually reducing hit detection efficiency~\cite{KADYK1991436,VAVRA1986547}.
These effects are strongest in the innermost drift-cell layers, where particle fluxes and background rates are highest.
As a consequence, the degradation is dependent on the transverse distance from the beam axis, with pronounced inefficiencies at small radii and a gradual decrease toward the outer layers, which can be seen in \cref{fig:cdc_ageing}.

\subsubsection*{Disabled readout boards}
\begin{figure}
    \centering
     \includegraphics[width=0.45\linewidth]{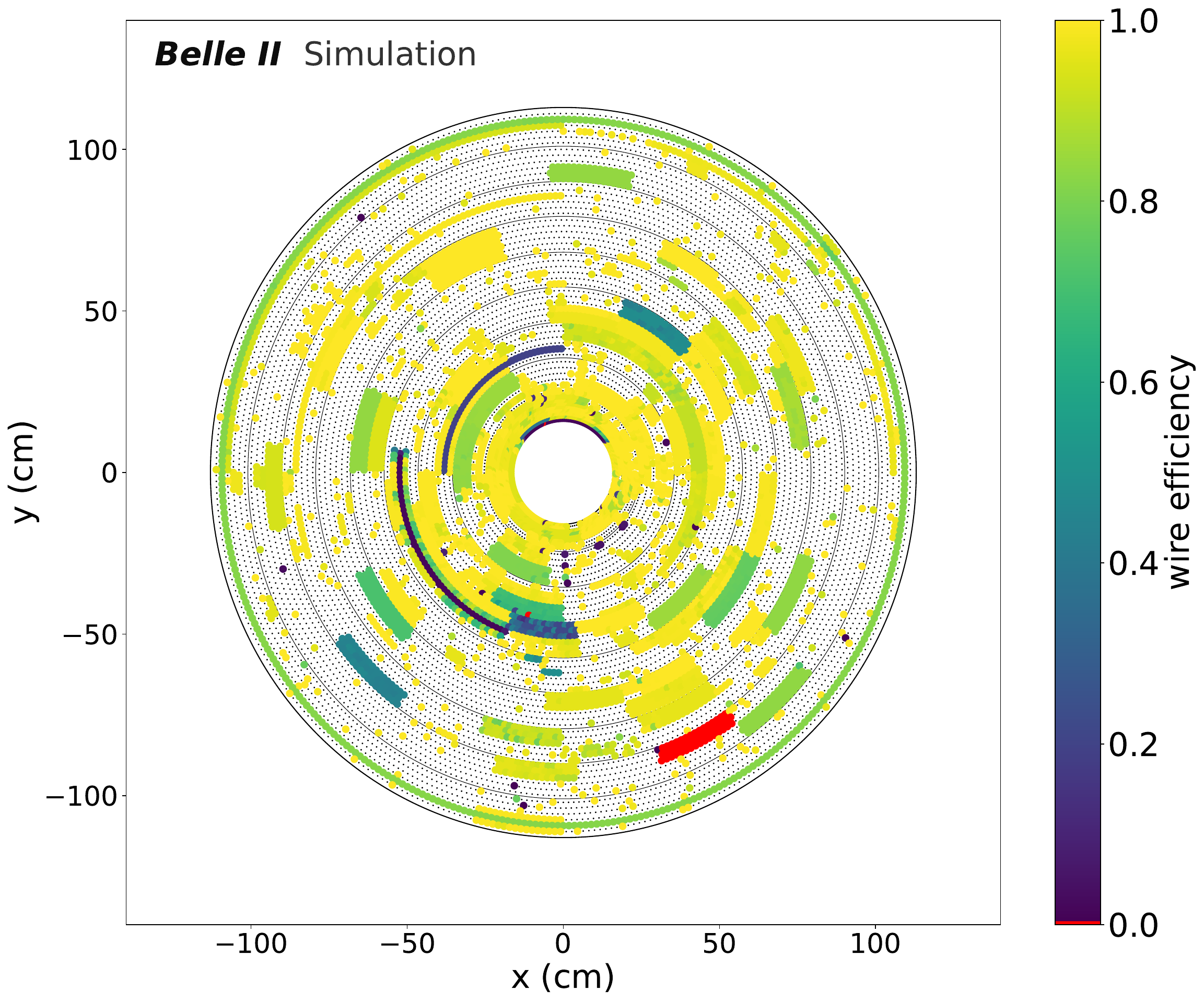}
    \caption{The distribution of the reduced wire efficiency, where the yellow, green and blue marked wires correspond to a decreased efficiency and regions in red correspond to inactive wires, weighted by integrated luminosity over the data taking from 2018 to 2022.}
    \label{fig:cdc_ageing}
\end{figure}

In addition to gradual ageing, radiation-induced failures or operational safety measures can necessitate the disabling of entire readout boards, showcased in  \cref{fig:cdc_ageing}.
This removes contiguous regions of the detector from readout, introducing localized gaps on the hit information on the scales from days to months.
Unlike wire ageing, these failures occur suddenly and can alter the spatial pattern of inefficiencies between events.
While this information can be incorporated in the offline reconstruction, the HLT must be able to operate reliably under these sudden changes, which further complicates track reconstruction.

\subsubsection*{Enlarged gap between SVD and CDC}
A particularly challenging consequence of \cdc degradation arises when the innermost CDC superlayers become inefficient or unavailable.
In this case, an increased radial gap of several centimeters occurs between the outermost Silicon Vertex Detector~(\svd) layers and the first CDC measurements.
This region normally provides geometric continuity between the two tracking detectors in the transverse plane, enabling reliable extrapolation of track candidates between the \svd and \cdc.
This enlarged gap significantly increases the associated uncertainties during the extrapolation.
As a result, track candidates may be reconstructed independently in the \svd and \cdc and cannot always be merged successfully.
In particular, \cdc-only tracks suffer from degraded longitudinal momentum resolution, reduced momentum resolution for low momentum particles and vertex resolution due to the absence of early inner-detector hits, leading to less precise extrapolation into the \svd.
This reduces the matching accuracy between track segments found by the \cdc and \svd separately and consequently increase the rate of duplicate (clone) tracks.

\subsubsection*{Impact on hit patterns and track reconstruction}
Taken together, these degradation mechanisms produce missing-hit patterns that violate key assumptions of conventional track reconstruction.
Tracks that previously yielded $\mathcal{O}(50)$ hits at transverse momenta above 400\,MeV/c may retain only a fraction of these measurements, with the remaining hits distributed irregularly along the trajectory.
In particular:
\begin{itemize}
    \item multiple consecutive hits can be missing in the inner CDC layers, which mostly affects the parameter resolution of low-momentum tracks,
    \item extended $\phi$-dependent inefficiencies arise from disabled readout boards,
    \item displaced tracks traverse fewer \svd layers and therefore rely more heavily on \cdc measurements, which have worse momentum and vertex resolution due to missing inner hits. 
\end{itemize}
Such hit patterns primarily violate the geometric continuity assumptions used in the \legendre \svd-to-\cdc CKF and the track quality estimator.
These effects are already observed during ongoing Belle~II data taking and are expected to intensify as SuperKEKB approaches its design luminosity.
Track finding can fail when gaps span entire superlayers for the axial layers, and hit attachment of stereo layers across extended inactive regions becomes ambiguous.
Strict continuity requirements suppress fake tracks but significantly reduce track finding efficiency, while relaxing these constraints leads to rapidly increasing fake and duplicate rates.
Adapting the staged reconstruction to these conditions would therefore require substantial algorithmic redesign.

\section{GNN Framework and Robustness to Detector Degradation}
\label{sec:gnn}

\begin{figure}
    \centering
    \includegraphics[width=0.9\linewidth]{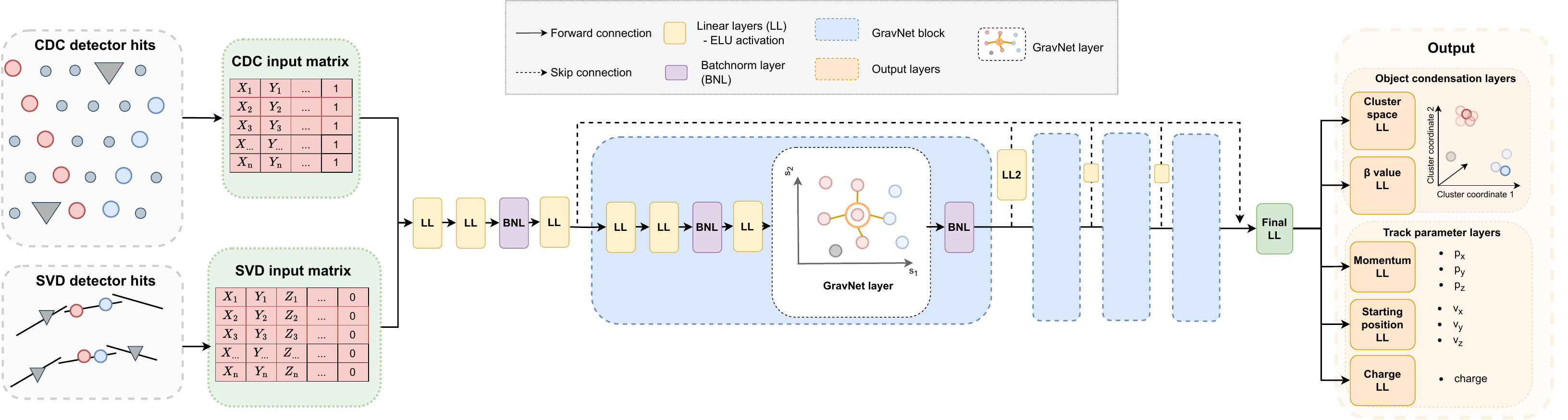}
    \caption{\bat model architecture, taken from \cite{acat}.}
    \label{fig:model}
\end{figure}

The GNN-based reconstruction used in this work extends on the CDC-only CDC AI Track Finder (\cat)~\cite{cat_paper}. 
This algorithm replaces the \cdc-only track finding in \cref{fig:baseline_tracking}, but keeps the staged approach with the \svd track finding.
Building on the successful GNN reconstruction for the \cdc, we also implemented a multi-modal GNN, using the combined \svd-\cdc hits as input in a single inference step. 
This new model is called the \belle AI Track finder (\bat), and we use the model and architecture defined in ~\cite{acat}.
In this new approach, the track finding is formulated as a single relational clustering problem over all detector hits in the event, rather than as a sequence of detector-dependent reconstruction steps.
All hits, independent of detector subsystem are treated as nodes in a single graph.
Hit-to-hit relations are established dynamically using the GravNet neighbourhood selection~\cite{Qasim:2019otl}.
Rather than relying on geometric proximity, neighbouring nodes are selected in a learned latent space.
This allows the model to learn to associate hits originating from the same charged particle even across large geometric gaps or in regions with highly non-uniform detector response.
Using this approach removes any explicit dependence on detector geometry, layer ordering, or hit continuity and makes the approach robust to irregular hit patterns arising from wire ageing, disabled readout boards, and enlarged gaps between the \svd and the first active \cdc layers.
\\
For each hit, the network predicts coordinates in a final latent space used for clustering, a condensation score, and initial track parameters.
The object condensation loss function~\cite{Kieseler:2020wcq} combines attractive and repulsive terms to form compact clusters in this final latent space containing hits from the same particles while suppressing noise hits.
The track candidates are then identified using the post-processing defined in~\cite{cat_paper} from the clusters in the cluster space.
Importantly, no loss term enforces hit continuity, regular spacing, or detector-specific ordering.

\section{Training Strategy for Degraded Detector Conditions}
\label{sec:training}

\begin{figure}
    \centering
    \subfloat[]{\includegraphics[width=0.488\linewidth]{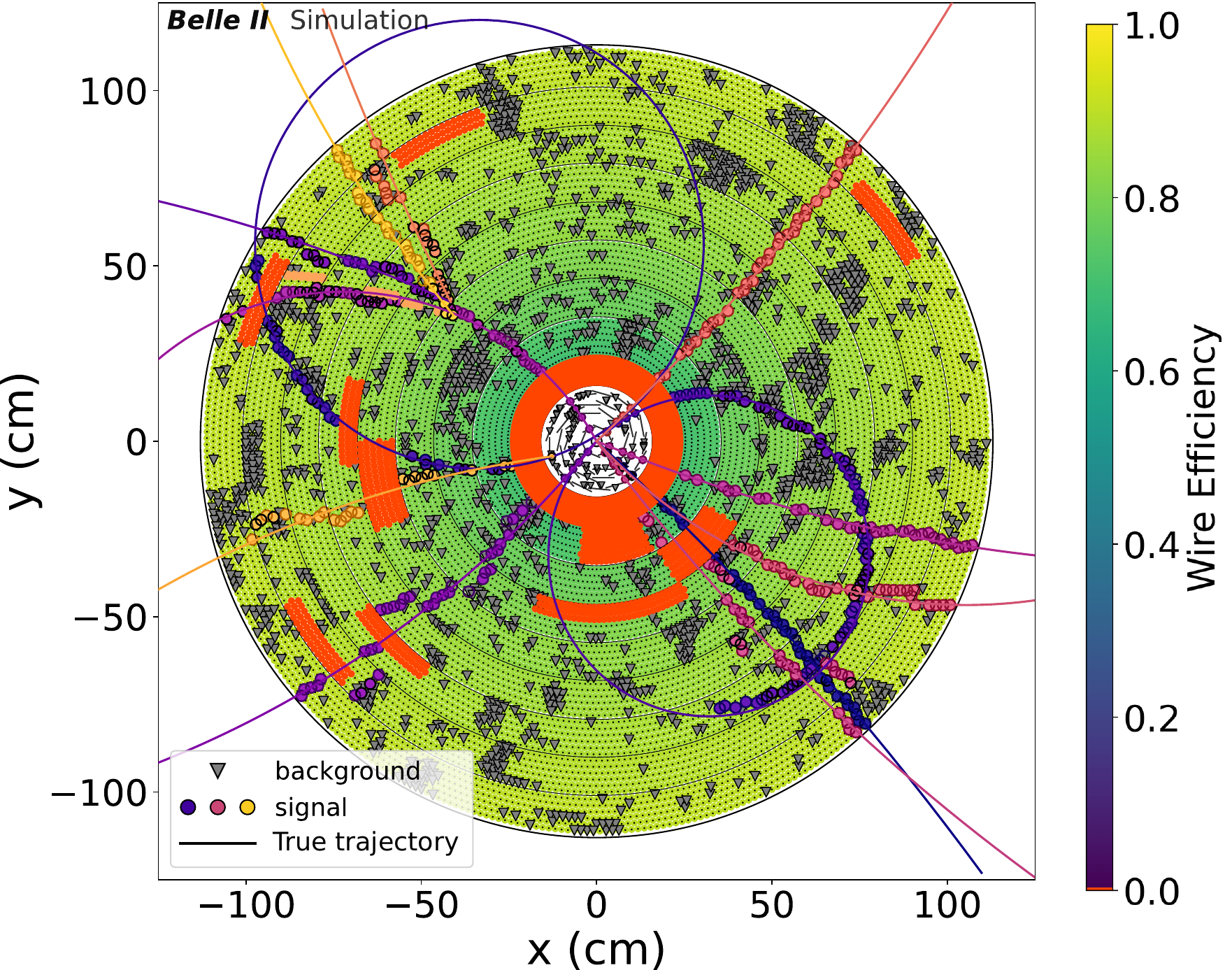}}
    \qquad
    \subfloat[]{\includegraphics[width=0.412\linewidth]{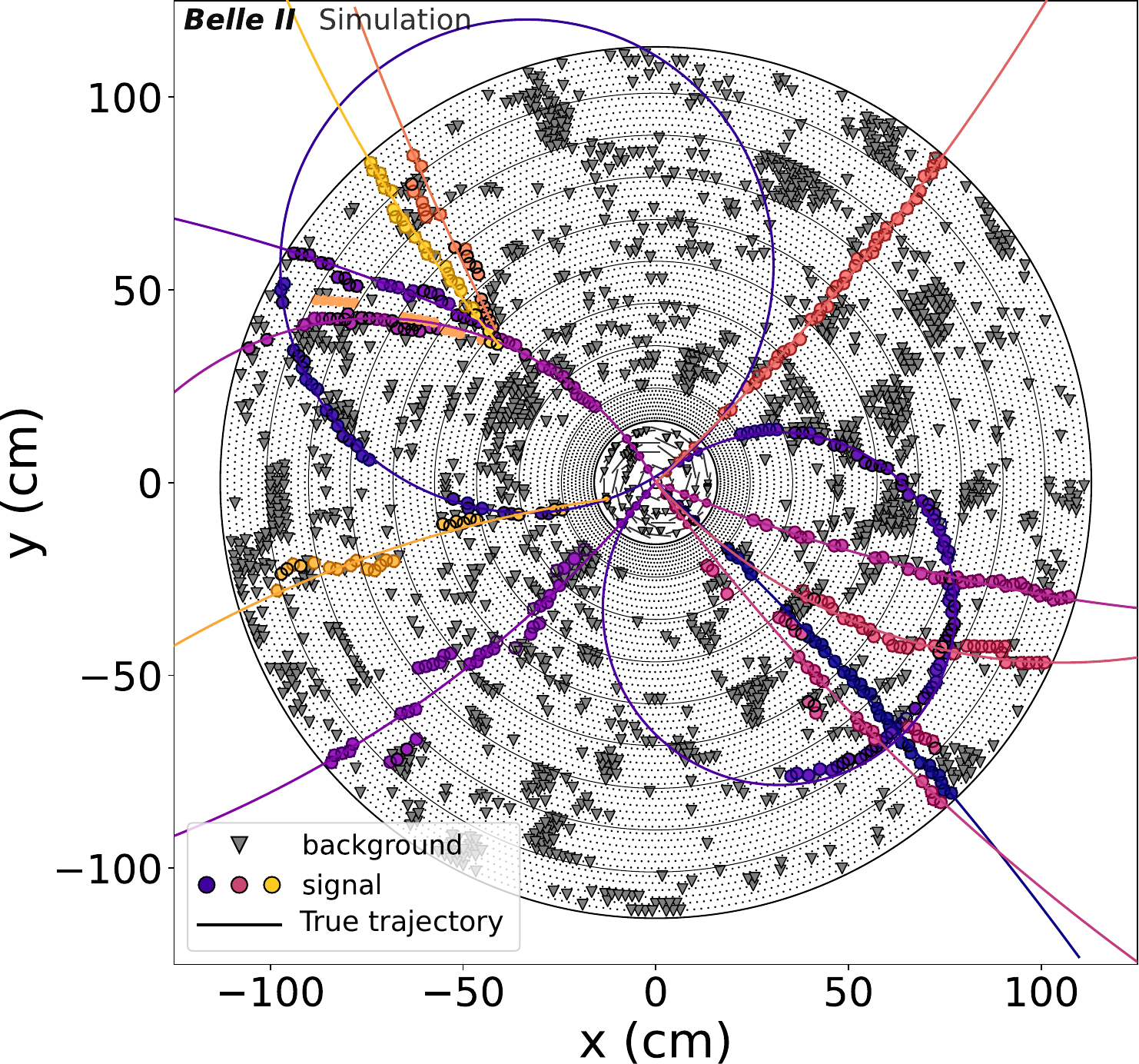}}
    \caption{Example event displays illustrating the current \cdc degradation (C=0.35, see \cref{eq:layereff} for details). Coloured markers indicate signal hits, while grey markers denote beam background. Figure (a) shows reduced wire efficiencies and inactive regions together with the resulting gaps, whereas figure (b) presents only the gaps for better visibility.}
    \label{fig:wiremap_event}
\end{figure}

Robustness to detector ageing is achieved through data-driven retraining rather than architectural modifications.
We simulate training samples that include realistic \cdc degradation scenarios derived from accumulated charge measurements and observed hardware failures, which will be explained in detail in the following.
\\
Following~\cite{lea_thesis}, the wire detection efficiency in layer $i$ is parametrised, in a simplified form, as
\begin{equation}
\label{eq:layereff}
    \layereff = 1 - C \cdot \frac{N_{0}}{N_{i}} \cdot \frac{L_{0}}{L_{i}},
\end{equation}
where $N_i$ and $L_i$ denote the number and length of wires in the $i$-th layer, respectively, and the factor $C$ controls the overall detector ageing, where a value of $C=1$ corresponds to a wire efficiency of 0 on the first layer.
A value of $C=0.35$ approximately reproduces the current \cdc conditions, yielding the strongest inefficiencies in the innermost layers, as shown in \cref{fig:wiremap_event}.
Wires with reduced efficiency register a hit only with the corresponding efficiency probability, resulting in randomly missing hits.
\\
In addition, disabled readout boards are randomly simulated  in spatially non-uniform patterns to reflect realistic failure modes during the design target instantaneous luminosity operation.
Scenarios with consecutive disabled boards and complete loss of the first superlayer are explicitly included.
Beam-induced background from simulation and recorded data~\cite{Liptak:2021tog,Natochii:2022vcs} is overlaid to reproduce realistic operating conditions.
\\
The training dataset is defined in detail in~\cite{cat_paper} and contains prompt muons, as well as muons up to 100\,cm displaced from the interaction point, with momenta between 50\,MeV/c to 6\,GeV/c and angular distributions covering the full tracking acceptance, simulated using the \belle Analysis Software Framework~\texttt{basf2}~\cite{basf21, basf22}.
Low-momentum muons below 400\,MeV/c and variable event multiplicities up to 15 particles per event are included to ensure sensitivity to the most challenging event topologies of high track multiplicity and low transverse momentum..
\\
The \bat architecture and inference procedure remain unchanged with respect to what was presented in \cite{acat}.
Previously trained \bat models are fine-tuned on degraded datasets, leading to fast convergence within 100 additional epochs instead of the required 800 epochs for retraining from the start and demonstrating that \cdc ageing primarily induces a domain shift in the hit patterns rather than a fundamentally new reconstruction problem.
\\
This training strategy supports two operational modes.
First, for track reconstruction on the HLT, training on a broad mixture of degradation patterns enables stable performance under rapidly changing detector conditions.
For this mixture, degraded readout boards and wire inefficiencies are simulated independently on an event-by-event basis, using one of up to 60 random configurations with 5 to 12 disabled readout boards and wire inefficiencies with factor $C=[0,0.25,0.5]$.
Second, for offline reconstruction, data are reprocessed with the measured detector degradation, and the model can be fine-tuned accordingly to improve efficiency and resolution.~\cite{lea_thesis}.
we focus exclusively on the second mode. 
The results presented here correspond to offline reconstruction using a single measured detector configuration, shown in \cref{fig:wiremap_event}(a), with $C=0.35$, a disabled first superlayer, and 14 disabled readout boards, including three consecutive combinations.

\section{Tracking Performance Under CDC Degradation}
\label{sec:performance}

\begin{table}[b]
    \centering
    \caption{Average track efficiency and track purity for displaced muons over the full detector region including statistical uncertainties under degraded CDC conditions for the multi-stage reconstruction of \legendre and \cat, and the one-shot approach of the \bat.
    The nominal results of the unaffected \cdc from \cite{acat} are shown for reference.}
    \label{tab:degraded}
    \begin{tabular}{lcc}
        \hline
        \hline
        \textbf{Algorithm} & \textbf{Track efficiency} & \textbf{Track purity} \\
        \hline
        \legendre (nominal) & $0.4804 \pm 0.0010$ & $0.9218 \pm 0.0006$ \\
        \hspace{0.5cm} \legendre (degraded CDC) & $0.3618 \pm 0.0012$ & $0.9018 \pm 0.0010$ \\
        \cat (nominal) & $0.6848 \pm 0.0008$ & $0.9380 \pm 0.0005$ \\
        
         \hspace{0.5cm} \cat (degraded CDC) & $0.5918 \pm 0.0014$ & $0.9131 \pm 0.0010$ \\
        \bat (nominal) & $0.7469 \pm 0.0007$ & $0.9763 \pm 0.0004$ \\
        \hspace{0.5cm} \bat (degraded CDC) & $0.6256 \pm 0.0012$ & $0.9337 \pm 0.0007$\\
          \hspace{0.5cm} \textbf{retrained} \bat (degraded CDC)  & $0.6417 \pm 0.0010$ & $0.9613 \pm 0.0010$ \\
        
        \hline
        \hline
    \end{tabular}
\end{table}
\begin{figure}
    \centering
    \subfloat[]{\includegraphics[width=0.455\linewidth]{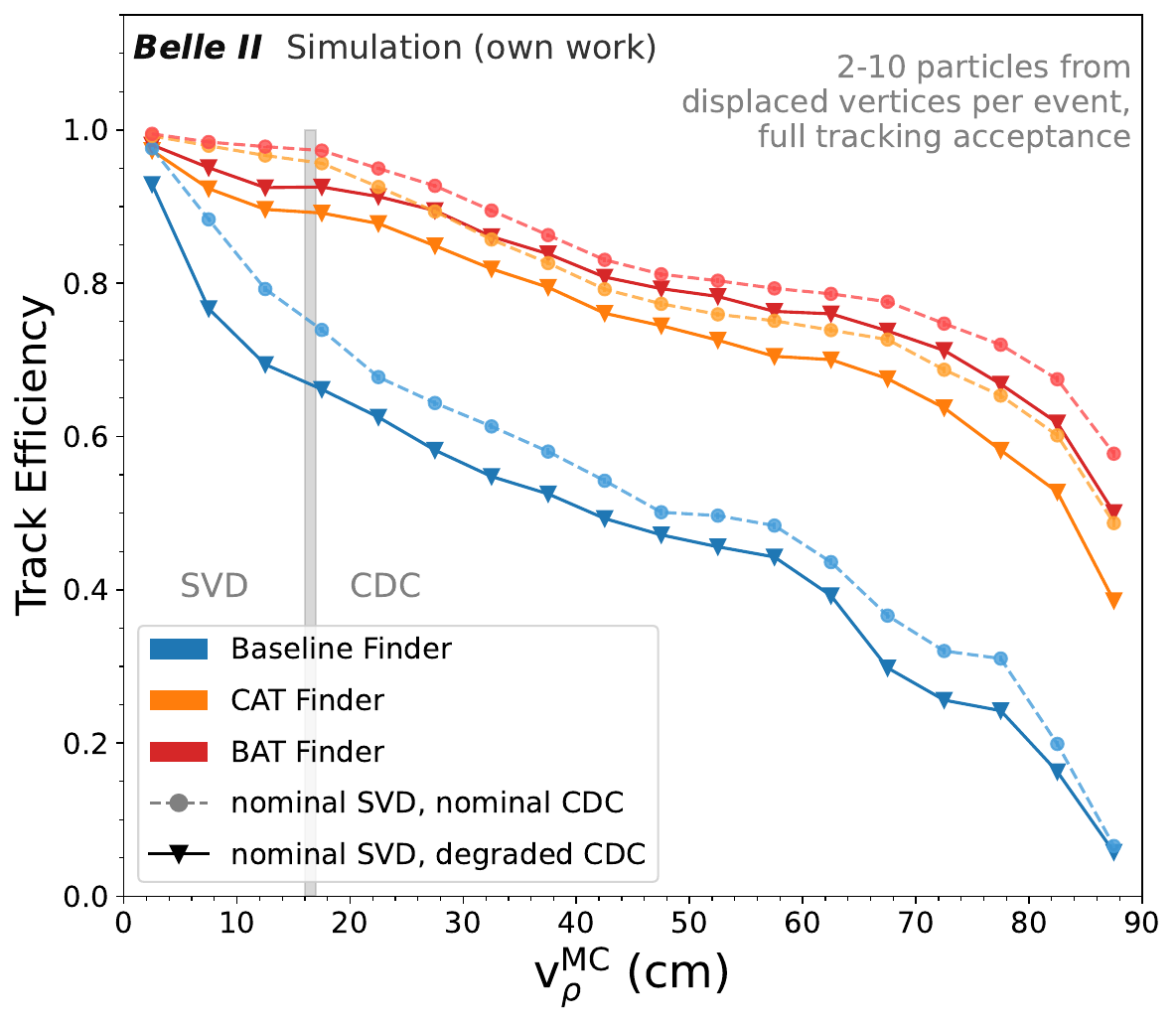}}
    \qquad 
    \subfloat[]{\includegraphics[width=0.445\linewidth]{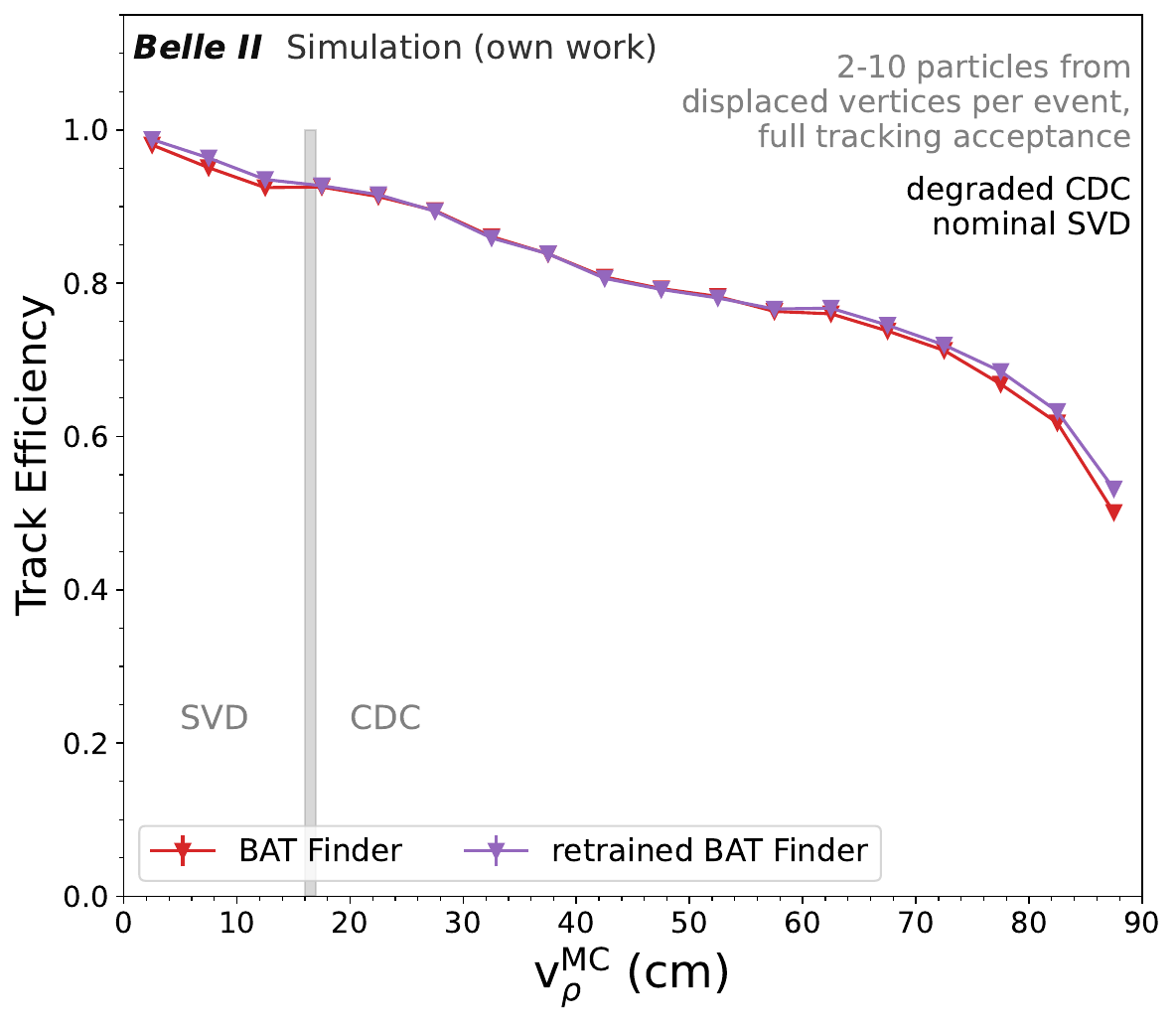}}
    \caption{Track efficiency as a function of transverse displacement $v_{\rho}^{\mathrm{MC}}$ for the  \legendre in blue, \cat in orange and retrained \bat in red. 
    The grey vertical line marks the transition between the SVD and the CDC, corresponding to the outer radius of the SVD.
    For displacements beyond this radius, tracks originate inside the CDC and no longer traverse the SVD.
    Here, (a) shows the track efficiency with the nominal SVD and CDC detector as a dotted line, taken as reference from \cite{acat}, and for the degraded CDC detector map shown in \cref{fig:wiremap_event} as solid line.
    The \bat (red) and the retrained \bat (purple) are shown in (b) for the degraded CDC.}
    \label{fig:efficiency}
\end{figure}

The performance evaluation focuses on opposite charged particles from displaced vertices up to a uniform displacement of 100\,cm, that are reconstructed under the realistic \cdc degradation scenario, as shown in \cref{fig:wiremap_event}.
Track efficiency is defined as $N_\mathrm{reco,matched}/N_\mathrm{true}$, where $N_\mathrm{reco,matched}$ is the number of  tracks matched to unique true particles and $N_\mathrm{true}$ is the number of charged particles producing at least one hit in the SVD or CDC~\cite{BelleIITrackingGroup:2020hpx,cat_paper}.  
Track purity is defined as $N_\mathrm{reco, matched}/N_\mathrm{reco, total}$, where $N_\mathrm{reco, total}$ is the total number of reconstructed tracks
~\cite{BelleIITrackingGroup:2020hpx,cat_paper}.
We compare three tracking approaches: the staged \legendre, the staged \cat, in which the GNN-based method replaces the CDC-only track finding while retaining the \belle SVD tracking, and the \bat, which performs a single inference combining both SVD and CDC hits.
Additionally, we include a retrained version of the \bat on the degraded conditions in the comparison.
The averaged track efficiency and track purity for \legendre, \cat, \bat, and the retrained \bat for muons from the uniformly displaced vertices are summarized in Table~\ref{tab:degraded}.
The comparison between nominal and degraded \cdc conditions highlights clear differences in robustness.\\
For the staged \legendre reconstruction, detector degradation results in a large relative track efficiency loss exceeding 24\%, together with a noticeable reduction in track purity.
This reflects the strong reliance on continuous hit patterns and intact inner-layer coverage.
The purity decreases further because degraded momentum and vertex resolution worsen the \svd-\cdc matching, increasing the rate of duplicate tracks.
\\
The \cat achieves a higher track efficiency than the \legendre under degraded conditions (59.2\% versus 36.2\%) and shows improved robustness, with a relative track efficiency loss of about 13.6\% when transitioning from nominal to degraded \cdc conditions.
However, due to the staged \cdc to \svd to \cdc reconstruction sequence, track segments from the same particle can be reconstructed independently in the \cdc and \svd in the presence of extended gaps.
Reduced resolution then prevents reliable merging of these segments, leading to a decrease in track purity.
Improving this behaviour would require structural modifications similar to those needed for the \legendre reconstruction.
\\
The \bat exhibits a slightly higher relative track efficiency loss of approximately 16\% without retraining, while achieving the highest overall track efficiency and purity. 
Furthermore, the relative decrease in track purity is about twice as large for the \bat without retraining compared to the staged methods.
This is expected, as the \bat performs combined SVD and CDC track reconstruction in a single inference step.
The enlarged radial gap between the \svd and \cdc is not represented in the training dataset for nominal conditions. As a result, hits from a single particle trajectory can be split into two segments, which the model may misidentify as separate particles, increasing the duplicate-track rate.
Retraining the \bat on the specific degraded detector conditions significantly reduces the duplicate rate and restores the track purity from $93.4\%$ to $96.1\%$.
At the same time, retraining also improves the track efficiency, increasing it from $62.6\%$ to $64.2\%$.
\\
Figure~\ref{fig:efficiency} shows the track efficiency as a function of transverse displacement $v_{\rho}^{\mathrm{MC}}$ under nominal \cdc conditions and degraded conditions~(a) for the three tracking algorithms without retraining, and the retrained \bat on degraded \cdc conditions in~(b).
For displacements beyond approximately 16\,cm, particles originate outside the \svd and traverse only the \cdc, thus the reconstruction relies exclusively on \cdc hits.
Here, both the \cat and the \bat consistently outperform the \legendre.
The additional improvement of the \bat over the \cat arises from the inclusion of enhanced \cdc hit information, as discussed in~\cite{acat}.
\\
At smaller displacements, tracks benefit from additional \svd hits.
In this region, both the \cat and the \bat exhibit greater robustness than the \legendre, showing a smaller track efficiency loss when comparing nominal and degraded \cdc conditions.
This behaviour is particularly relevant for all prompt particles as well as particles with a decay vertex within the \svd, such as decay products from a $K^0_S$ with a lifetime of $c\tau=$2.68\,cm~\cite{ParticleDataGroup:2022pth}.
Within the \svd region, the track efficiency loss of the \legendre remains significant, while the efficiencies of the \cat and \bat saturate once tracks traverse only about half of the \svd layers.
The reduced number of hits and the enlarged radial gap reduce the reconstruction effectively to the \cdc-only case, preventing reliable attachment of \svd hits.
Retraining the \bat recovers the track efficiency in this region by incorporating \svd information more effectively within the combined SVD and CDC track finding.
This improvement is particularly relevant for low-momentum tracks and particles exiting through the endcaps, which do not reach the outer, less affected \cdc layers.
A second region of improvement from retraining is observed at larger displacements, where the number of available \cdc layers decreases.
In this region, retraining enables the \bat to better handle sparse hit patterns, leading to improved track efficiency compared to the non-retrained model.
\\
The improved robustness of the retrained \bat to degrading detector conditions arises from the joint clustering of \svd and \cdc hits within a single graph.
By avoiding intermediate detector-to-detector matching steps, the unified approach prevents the formation of independent track segments across enlarged gaps between the \svd and the first active \cdc layers.
In contrast, staged reconstruction accumulates inefficiencies through repeated matching failures, leading to increased duplicate rates and reduced track purity.
\\
Realistic \cdc ageing leads to sparse and irregular hit patterns due to missing layers and localized inactive regions.
Although detector degradation reduces hit multiplicity and introduces spatially non-uniform inefficiencies, the correlations between hits originating from the same charged particle remain largely intact.
The GNN-based reconstruction is inherently robust to such conditions, as tracks are identified as clusters in a learned latent space without relying on explicit detector geometry, layer ordering, or hit continuity.
This enables stable track reconstruction across extended gaps and missing measurements, maintaining a track purity of about 96\% for the retrained \bat.
Retraining on datasets that reflect the degraded \cdc conditions further improves both track efficiency and track purity without any modifications to the model architecture or inference procedure.
This provides a simple mechanism to adapt reconstruction performance as detector conditions evolve, in contrast to staged approaches that degrade abruptly.
Such behaviour is essential for sustained \cdc operation under continuous ageing and intermittent hardware failures.

\section{Conclusion}
\label{sec:conclusion}
We have presented a study of GNN-based track reconstruction under realistic \cdc degradation conditions.
Detector ageing mainly alters the observed hit patterns rather than introducing a fundamentally new reconstruction problem.
By retraining an existing GNN-based track finder on degraded detector samples, without changing the model architecture or inference procedure, a large fraction of the lost track efficiency is recovered.
For muons from uniformly displaced vertices up to 100\,cm, the \bat achieves a track efficiency of 62.6\%, compared to 36.2\% for the \legendre, while maintaining a higher track purity of 96\% compared to 90\%.
Retraining provides a comparibly fast and automatable way to adapt the reconstruction to changing detector conditions.
This makes the GNN-based approach well-suited for continued operation as the \cdc performance degrades over time.


\Acknowledgements
We are grateful to the members of the \belle tracking group for discussion and feedback. We thank Carsten Niebuhr for his input on the modelling of the \cdc degradation due to accumulated charge.



\end{document}